\documentclass[a4paper,11pt]{article}
\usepackage{pos}
\usepackage[running]{lineno}
\usepackage[utf8]{inputenc}

\title{Multi-messenger and real-time astrophysics with the Baikal-GVD telescope}
 \ShortTitle{Multi-messenger fast analysis of data at Baikal-GVD}
\author[a]{V.A.~Allakhverdyan}
\author[b]{A.D.~Avrorin}
\author[b]{A.V.~Avrorin}
\author[b]{V.M.~Aynutdinov}
\author[c]{R.~Bannasch}
\author[d]{Z.~Barda\v{c}ov\'{a}}
\author[a]{I.A.~Belolaptikov}
\author[a]{I.V.~Borina}
\author[a,1]{V.B.~Brudanin}
\author[e]{N.M.~Budnev}
\author[a]{V.Y.~Dik}
\author[b]{G.V.~Domogatsky}
\author[b]{A.A.~Doroshenko}
\author[a,d]{R.~Dvornick\'{y}}
\author[e]{A.N.~Dyachok}
\author[b]{Zh.-A.M.~Dzhilkibaev}
\author[d]{E.~Eckerov\'{a}}
\author[a]{T.V.~Elzhov}
\author[f]{L.~Fajt}
\author[g,1]{S.V.~Fialkovski}
\author[e]{A.R.~Gafarov}
\author[b]{K.V.~Golubkov}
\author[a]{N.S.~Gorshkov}
\author[e]{T.I.~Gress}
\author[a]{M.S.~Katulin}
\author[c]{K.G.~Kebkal}
\author[c]{O.G.~Kebkal}
\author[a]{E.V.~Khramov}
\author[a]{M.M.~Kolbin}
\author[a]{K.V.~Konischev}
\author[h]{K.A.~Kopa\'{n}ski}
\author[a]{A.V.~Korobchenko}
\author[b]{A.P.~Koshechkin}
\author[i]{V.A.~Kozhin}
\author[a]{M.V.~Kruglov}
\author[b]{M.K.~Kryukov}
\author[g]{V.F.~Kulepov}
\author[h]{Pa.~Malecki}
\author[a]{Y.M.~Malyshkin}
\author[b]{M.B.~Milenin}
\author[e]{R.R.~Mirgazov}
\author[a]{D.V.~Naumov}
\author[a]{V.~Nazari}
\author[h]{W.~Noga}
\author[b]{D.P.~Petukhov}
\author[a]{E.N.~Pliskovsky}
\author[j]{M.I.~Rozanov}
\author[a]{V.D.~Rushay}
\author[e]{E.V.~Ryabov}
\author[b]{G.B.~Safronov}
\author[a]{B.A.~Shaybonov}
\author[b]{M.D.~Shelepov}
\author[a,d,f]{F.~\v{S}imkovic}
\author[a]{A.E. Sirenko}
\author[i]{A.V.~Skurikhin}
\author[a]{A.G.~Solovjev}
\author[a]{M.N.~Sorokovikov}
\author[f]{I.~\v{S}tekl}
\author[b]{A.P.~Stromakov}
\author[a]{E.O.~Sushenok}
\author*[b]{O.V.~Suvorova}
\author[e]{V.A.~Tabolenko}
\author[e]{B.A.~Tarashansky}
\author[a]{Y.V.~Yablokova}
\author[c]{S.A.~Yakovlev}
\author[b]{D.N.~Zaborov}

\affiliation[a]{Joint Institute for Nuclear Research, Dubna, Russia}
\affiliation[b]{Institute for Nuclear Research, Russian Academy of Sciences, Moscow, Russia}
\affiliation[c]{EvoLogics GmbH, Berlin, Germany}
\affiliation[d]{Comenius University, Bratislava, Slovakia}
\affiliation[e]{Irkutsk State University, Irkutsk, Russia}
\affiliation[f]{Czech Technical University in Prague, Prague, Czech Republic}
\affiliation[g]{Nizhny Novgorod State Technical University, Nizhny Novgorod, Russia}
\affiliation[h]{Institute of Nuclear Physics of Polish Academy of Sciences (IFJ~PAN), Krak\'{o}w, Poland}
\affiliation[i]{Skobeltsyn Institute of Nuclear Physics, Moscow State University, Moscow, Russia}
\affiliation[j]{St.~Petersburg State Marine Technical University, St.Petersburg, Russia}

\note{Deceased.}

\emailAdd{suvorova@inr.ru}
\abstract{The Baikal-GVD deep underwater neutrino experiment participates in the international multi-messenger program on discovering the astrophysical sources of high energy fluxes of cosmic particles, while being at the stage of deployment with a gradual increase of its effective volume to the scale of a cubic kilometer. In April 2021 the effective volume of the detector has been reached 0.4 km$^3$ for cascade events with energy above 100 TeV generated by neutrino interactions in Lake Baikal. The alarm system in real-time monitoring of the celestial sphere was launched at the beginning of 2021, that allows to form the alerts of two ranks like "muon neutrino" and "VHE cascade". Recent results of fast follow-up searches for coincidences of Baikal-GVD high energy cascades with ANTARES/TAToO high energy neutrino alerts and IceCube GCN messages will be presented, as well as preliminary results of searches for high energy neutrinos in coincidence with the magnetar SGR 1935+2154 activity in period of radio and gamma burst in 2020.}

\FullConference{37$^{\rm{th}}$ International Cosmic Ray Conference (ICRC 2021)\\
		July 12th -- 23rd, 2021\\
		Online -- Berlin, Germany}

\begin{document}
\maketitle
\

\section{Introduction}

A real-time astrophysics is a nowadays tool in discovery of astrophysical sources of high energy neutrinos. 
Correlation analysis of alarm messages of a signal observation in electromagnetic frequencies with follow up of high energy neutrino events as well a neutrino alarms to astronomy fellow are considered as a way for identification of neutrino sources and the particle generation mechanisms at high energy. Do neutrinos correlated with known multimessenger signals? An answer could be in development of wide network of neutrino alerts and joint analysis of data from different experiments. It is well-know that the first neutrino alert IC170922A with an energy above 300 TeV recorded by the IceCube telescope toward the blazar TXS 0506+05 in the period of its activity in 2017~\cite{IC170922A} motivated multiwavelength observations of this blazar~\cite{multi_TXS2018}. The recent estimates of neutrino fluxes from astrophysical objects of radio loud blazars in class of Active Galactic Nuclei (AGN) claims that its part does not exceeds 30\% in diffuse neutrino flux~\cite{CT_2020_1}. Other class of  potential sources with non-thermal emission of high energy neutrino  are transients in such catastrophical phenomena as a tidal disruption events (TDE), compact object mergers followed by gravitational waves, core-collapse supernova and gamma-ray bursts. Follow-up of transients is quite complicated due to unknown burst time  of the physical processes.  

Really observed sources of multiwave signals are magnetars being the neutron stars powered by magnetic energy dissipation. Often they are discovered as Soft Gamma-Ray Repeaters (SGRs). The INTEGRAL discovered a new period of activity of the repeater SGR 1935+2154~\cite{INTEGRAL}, which performed 
tens of bursts in a short time interval in April 27-28 of 2020, first time confirmed association between an Fast Radio Bursts (FRB) and a high-energy bursting source.
The Baikal-GVD followed up of the INTEGRAL busrt alert in direction of a possible association with the supernova remnant G57.2+0.8.

At the begining of 2021, the alert system of the Baikal-GVD detector~\cite{BaikalNIM2014} was resulted in fast regime of event reconstruction in runtime data collecting, 
which is of 12 hours. That allows to form internal alerts of two ranks of neutrino-like events: “track of muon neutrino” and “VHE cascade”. In further development of the data collecting and processing (B.Shaybonov, this conference, PoS(ICRC2021)1040), we are going to form neutrino alarms to other communities. 
%(Id487 of presentation by B.Shaybonov) 
In fast regime of data performance we began the follow-up analysis of the IceCube muon neutrino alerts with energy above a hundred TeV, catching GCN/AMON messages since August 2020, as will be presented below.
Earlier, since December  2018, the Baikal-GVD collaboration triggered program of follow-up of muon neutrino alerts of the automated TAToO system~\cite{tatoo_2011} 
realized at the ANTARES telescope. No coincidences between the ANTARES alerts and the GVD data were found so far in follow-up studies, while a preliminary summary is obtained and discussed.

\section{Internal alert performance}
At present, Baikal-GVD is a largest operating neutrino telescope at the North latitudes with current effective volume of about 0.4 km$^3$ for electron neutrinos detection in energy range of hundred TeV. As planned, it will be twice increased in next years (I.Belolaptikov, V.Aynutdinov, this conference, PoS(ICRC2021)002 and PoS(ICRC2021)1066).
%(Id880 and Id747 of presentations by I.Belolaptikov, V.Aynutdinov). 
The real data production of the telescope is about 100 GB per cluster per day (B.Shaybonov, this conference, PoS(ICRC2021)1040).
%(Id487 of presentation by B.Shaybonov). 
Trasmission data stream is splitted into 250 MB files and transferred to the shore station through an optical fiber cable, stored there and then transmitted at 40 km through radio channel to Baikalsk (Irkutsk region) and then transferred through the Internet to storage in Dubna facility of the Joint Institute for Nuclear Research at 4300 km away from the GVD. To develope the online data processing, an essential progress has been achieved in fall 2020.  An internal high-energy neutrino alert is formed among upward going events as a result of the event reconstruction after the completion of a successive data transmission. Presently, the data processing mode, where the processing is started when obtaining all files of the data accumulation session  (runtime is 12 hours), has been completely realized and is being used. The session processing duration and, accordingly, the delay in forming an alert event with given characteristics takes in avearage 3--5 hours. There is a dependence of the delay on the background luminescence level of the lake and so far the number of files in the session. Rates of noise pulses vary between 20 and 100 kHz depending on season and depth. Under noise suppression by causality requirement $\pm$10 ns in the event reconstruction algorithms
%reconstruction the GVD optical modules hits the Cherenkov light detection of passing particles generated in the HE neutrino-nuclear interactions in Baikal water.
%there are realised two ranks of events in fast and general algorithms of reconstruction: 
there are two ranks of selection: track-like of HE muon and shower-like of VHE particles from cascade decays~\cite{Cscd},\cite{Trk}. The fast and general algorithms include the quality cuts on reconstructed parameters according to MC studies of their optimisations (Zh.Dzhilkibaev, G.Safronov, D.Zaborov, this conference, PoS(ICRC2021)1144, PoS(ICRC2021)1080, PoS(ICRC2021)1177).
%(see Id900, Id1011 and Id1449 of presentations by Zh.Dzhilkibaev, G.Safronov, D.Zaborov).\\
An accuracy of reconstructed angles for cascade events is about 2--4.5$^\circ$ as median values depending on cascade location and orientation. The energy resolution averaged by E$^{-2}$ spectrum of electron neutrinos 
is about 10--30$\%$ as well, about 90\% of these events are within range of 5 TeV$<$ E $<$ 10 PeV. An estimate of number of astrophysical events of E$^{-2.46}$ spectrum with energies above 100 TeV is about 0.4--0.6 per cluster per year and about 0.08 from atmospheric neutrinos (see PoS(ICRC2021)1144). Applied quality cuts reject events by number of hit OMs on string($>$ 7), by value of $\chi^2$ in reconstruction of vertex position and by likelihood value in reconstruction of energy and direction as well by product of probabilities of hits and no hits across correspondent optical modules. Cascades selected by the cuts have minimal energy of 1 TeV and are sampled for further studies. 
Last test of each selected cascade with energy of cascade E$_{sh}>$ 60 TeV includes quantified proof of the origin of hits and rejection of those confirmed by time residuals of incoming muon. In total, seven cascade events with E$_{sh}>$ 60 TeV have been selected as astrophycial candidates in data sample of 2019 and 2020 seasons (see PoS(ICRC2021)1144). Current procedure of reconstruction of muon trajectory is based on analysis of hits in a single cluster. For upward going muons the energy threshold is about 500 GeV. Fast algorithm rejects near horizon upward going events (i.e. zenith angles less than 120$^\circ$) to get high signal-to-noise ratio, while in off-line analysis reconstrusted tracks are tested in follow-up study. New algorithm for track reconstruction includes updated implementation of BDT differentiation and energy estimator (see PoS(ICRC2021)1080). Actual request is a reconstruction of multi clusters events related to a very high energy upward going muons. The left panel of Fig.1 shows an example of rates of multi clusters coincidences in the time window corresponding to the distance between clusters during a day in a season 2020.

%%%
%%%    Ris_1
%%%
\begin{figure}[!htb]
%\begin{center}
\includegraphics[width=0.45\textwidth,height=0.29\textwidth]{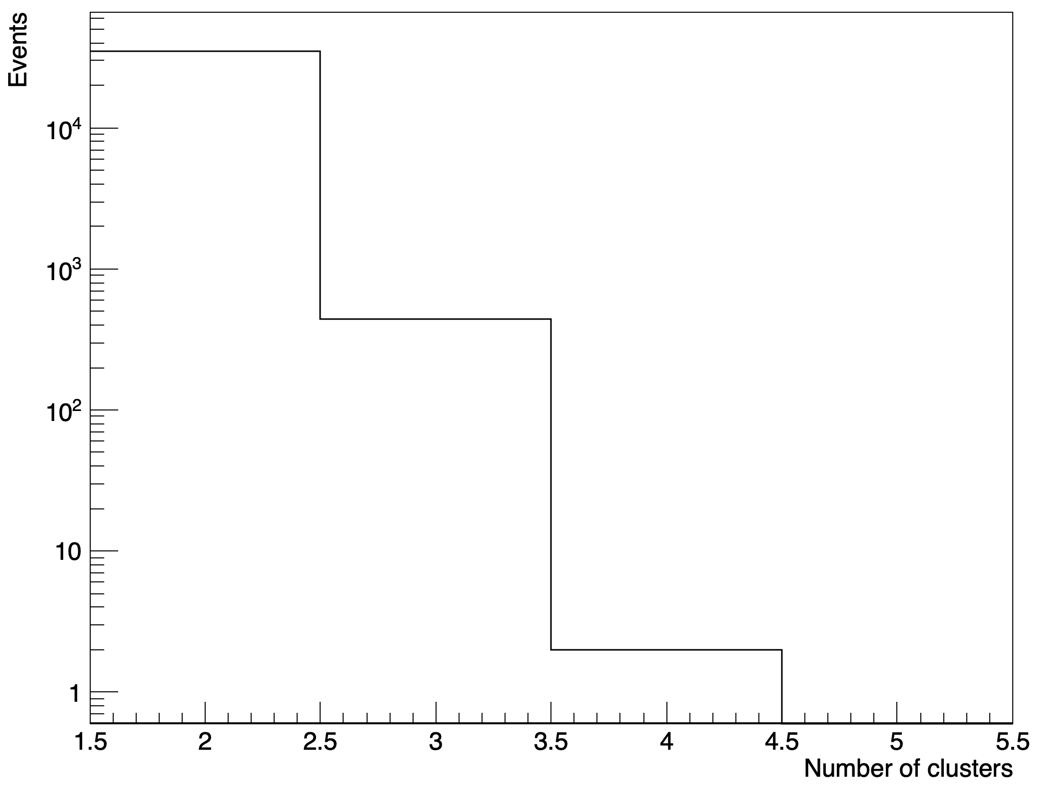}
\includegraphics[width=0.55\textwidth,height=0.35\textwidth]{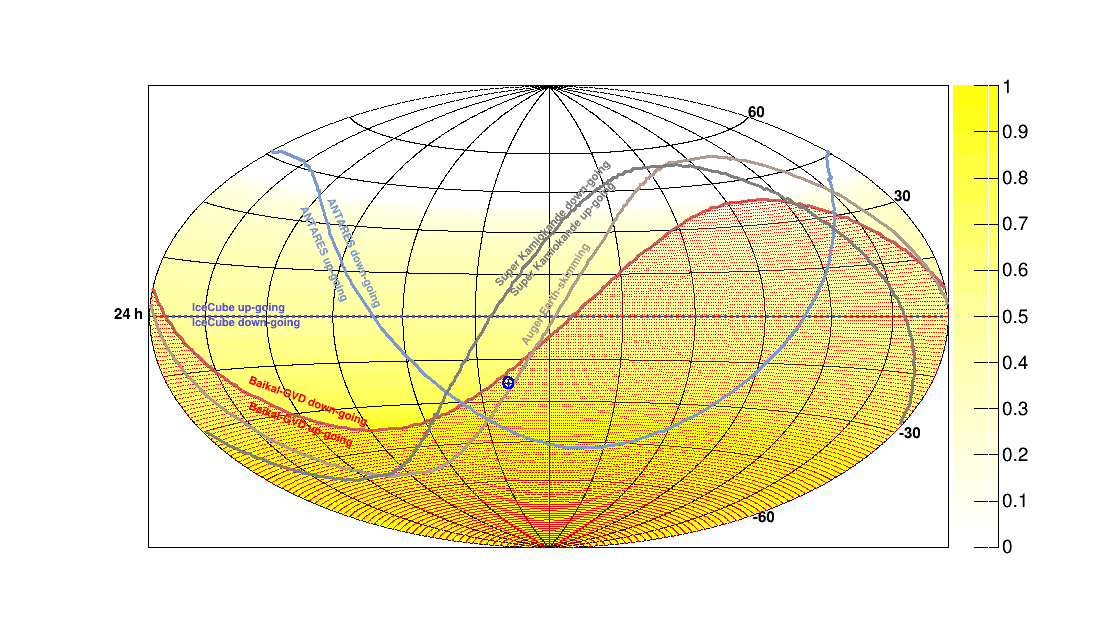}
%\end{center}
\caption{Left: Numbers of coincidences during a day of two or more GVD clusters within 6 $\mu$s according to the inter-cluster distance. 
Right: Local horizons of different experiments in arrival time of a gravitational wave alert GW170817A (blue circle). Shown are the Baikal-GVD (red line), ANTARES (blue) and IceCube (equator),
also Super Kamiokande (grey) and Pier Auger (brown). Scale of yellow color refers to sky visibility with Baikal-GVD upward going events. 
}
\label{fig-1}
\end{figure}
%%%
%%%    Ris_2
%%%
\begin{figure}[!htb]
%\begin{center}
\includegraphics[width=0.50\textwidth,height=0.35\textwidth]{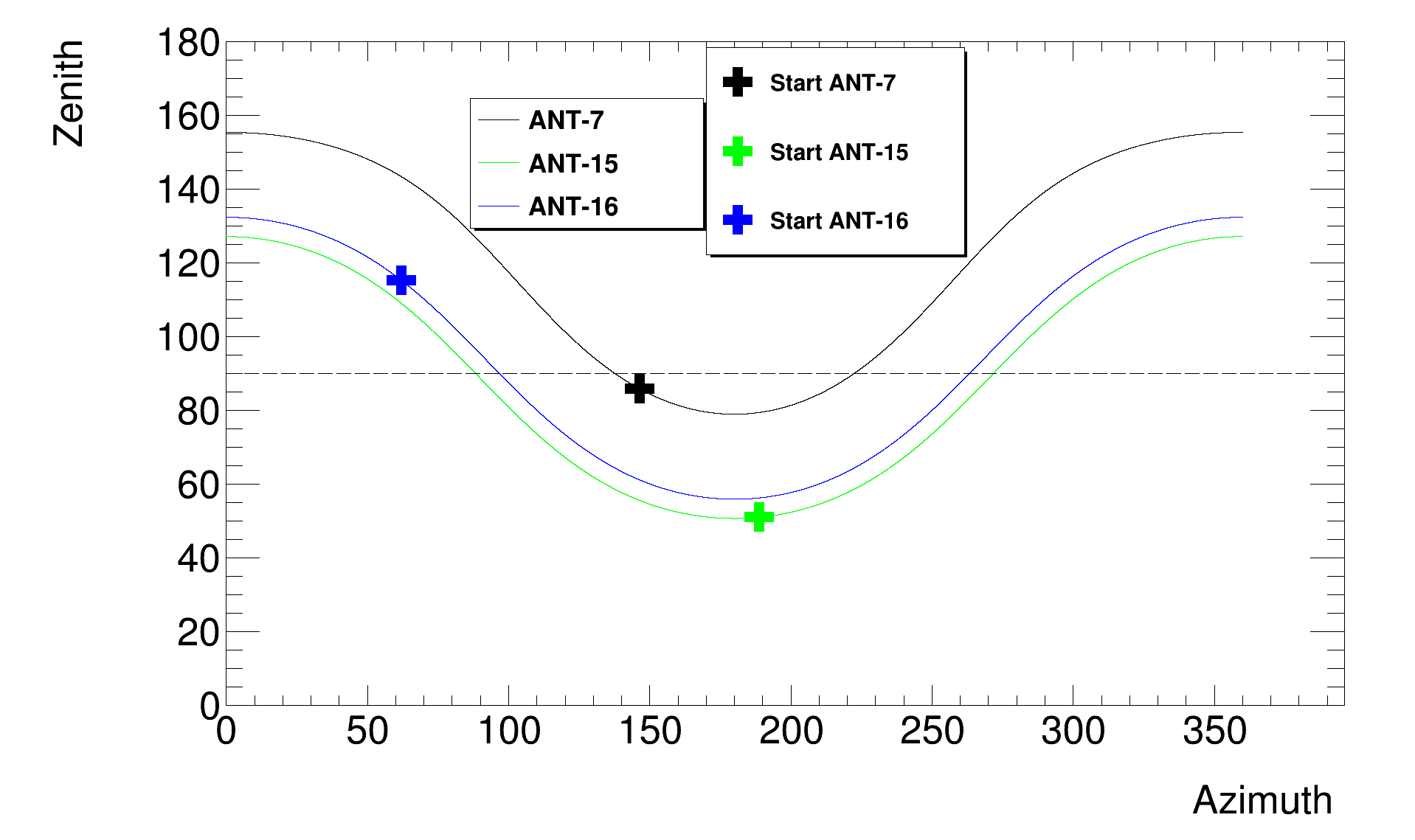}
\includegraphics[width=0.50\textwidth,height=0.35\textwidth]{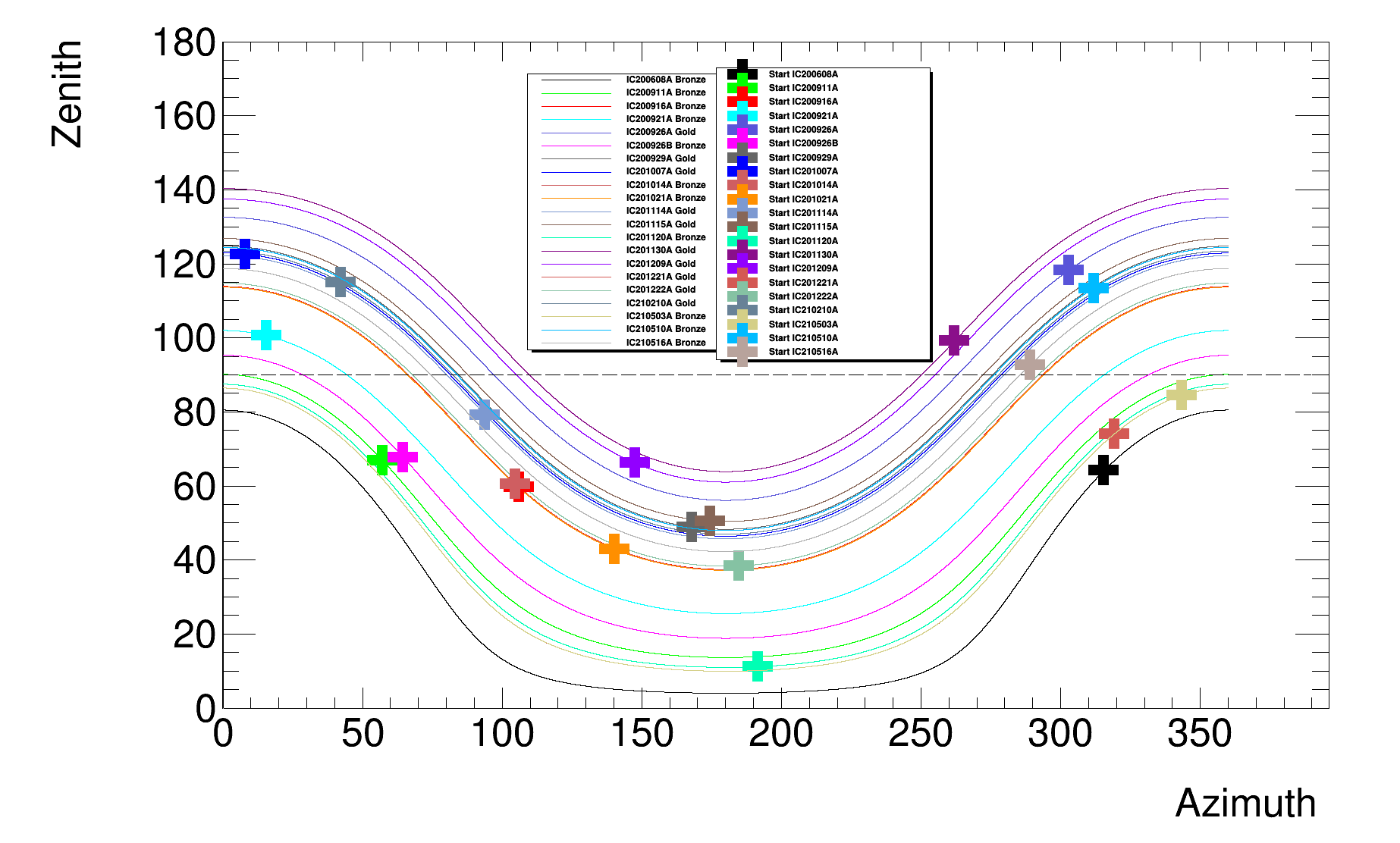}
%\end{center}
\caption{Diurnal trajectories of the fixed equatorial coordinates of neutrino alerts in Baikal-GVD horizontal coordinates. The crosses indicate alert locations.
Left: the ANTARES events selected for furter joint analysis due to GVD cascades follow-up. Right: The IceCube astrotracks in Fall 2020. See text for details.
}
\label{fig-2}
\end{figure}
\section{Follow-up of HE neutrino alerts}
The Baikal-GVD experimental data have been accumulated in a continuous exposure mode since 2015, allowing a prompt data analysis. A celestial-sphere monitoring 
program planned to be implemented in a real time soon.  
%Since 2015, the Baikal-GVD experiment is taking data in continuous mode, that allows for a real-time monitoring program for 
%the celestial sphere, while the telescope is build in a gradual  increase of its effective volume. 
%%%
%%%
%%%    Ris_3
%%%
\begin{figure}[!htb]
%\begin{center}                                                                            
\includegraphics[width=0.33\textwidth,height=0.35\textwidth]{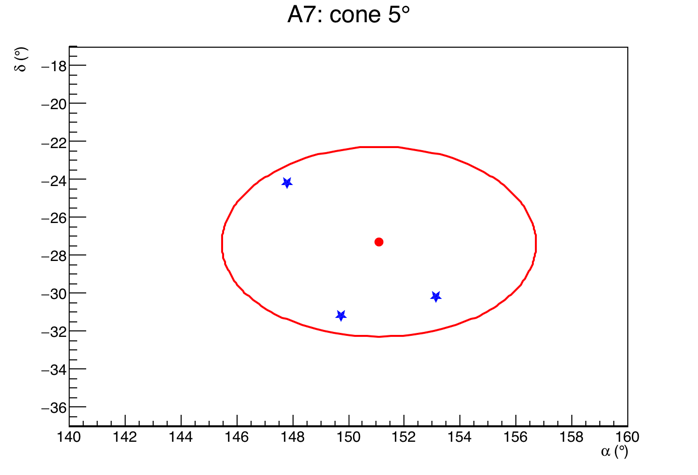}
\includegraphics[width=0.33\textwidth,height=0.35\textwidth]{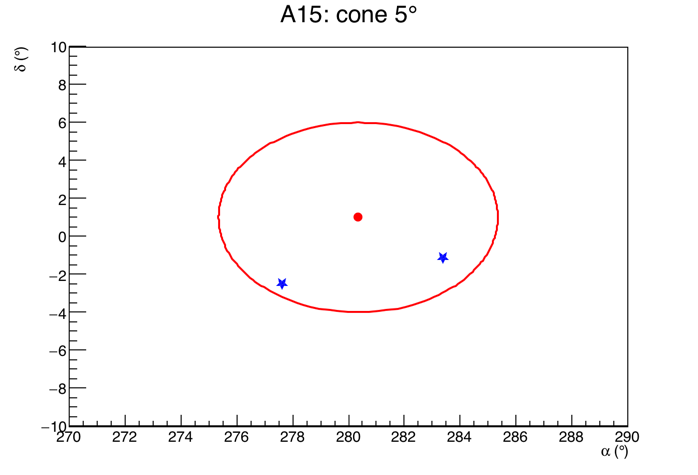}
\includegraphics[width=0.33\textwidth,height=0.35\textwidth]{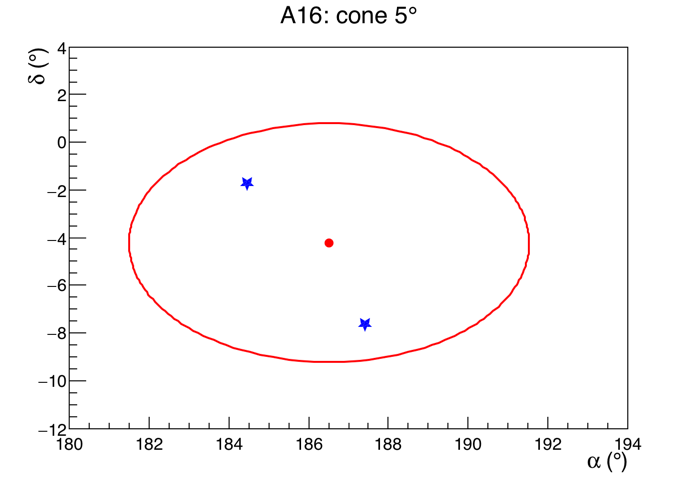}
%\end{center}
\caption{View in equtorial coordinates: circles of 5 degrees radius around the ANTARES alerts (red point in center) and the GVD repated cascades (blue stars). 
}
\label{fig-3}
\end{figure}
%%%
%%%
%%%
Presently, our first priority in searching for correlations in time and direction of the reconstructed GVD events with astrophysical alerts are high energy neutrino alerts i.e. 
alerts of the ANTARES and the IceCube neutrino telescope. All three detectors are complementary in simultanious observation of the sky and a duration scanning of the same astrophysical objects, which are considered as potential neutrino sources. Fig.1 (right) shows the lines of horizon for upward- and 
downward- going events at the telescopes in arrival time of multimessenger signal of a gravitational wave event GW170817A. The Baikal-GVD upper limits on neutrino fluence from 
the host galaxy NGC 4393  of two merging neutron stars are presented in~\cite{gvdGW}. 
The Baikal-GVD has received the ANTARES alerts of upwardgoing muons in a real-time since December 2018. In average, an accuracy of the ANTARES alert coordinates is up to 0.5$^{\circ}$  and energy of alerts is about few TeV. Following the alerts, we look for events on each cluster in time windows $\pm$500 sec, $\pm$1 h and $\pm$1 day around alerts inside a half-open cone according to the GVD angular resolutions. In total, we have received 46 alerts. No prompt coincidence in time and direction has been found. 
However, for three alerts in 2019 we have found repeated cascades in time window $\pm$ 1 day. The left panel of Fig.2 presents diurnal visibilities of these three alerts. Their zenith angles are less than 120$^\circ$ and thus no track-like events are in fast reconstruction. Fig.3 shows repeated cascades in a circle of 5${^\circ}$ around each alert in coordinates of declination and right ascension. For the alert A15 and A16 the energy of repeated cascades are about 3--4 TeV, while for the alert A7 the cascade energies are quite different: 2.9 TeV, 13.5 TeV and 158.0 TeV. We obtain the expected number of backgroud atmospheric muons by scrambling of real data of season 2019 and 2018. For total livetime of 690 days, there is an agreement between number of observed and expected events in a cone of 5$^\circ$ towards the source: 14/14.64 (A7); 32/37.23 (A15); 37/31.21 (A16). In time window of $\pm$ 1 day for coincidences of events in a cone of 5${^\circ}$ p-value to reject a hypothesis of background only events became 3 sigma. Further study of atmospheric muon bakground has been developed in a joint work of the ANTARES and GVD groups, that results in the upper limits on neutrino fluences as presented at the conference (S.A.Garre, PoS(ICRC2021)1121).
%(Id529) by S.A.Garre

% A7_Zenith= 79.5; 81.9(-3.2hours); 101.2(-23.2h); 84.9(+20.8h).
% 
% A15_Zenith= 51.1; 53.1(-0.64h); 82.2(+20.4h).
% 
% A16_Zenith= 115.1; 67.7(-18.7h); 60.5(-14.4h).
% 
% In summary, for "two "signal" events in +-days:
% 
% A7   p-val(2,0.090)=0.000846 / 3.1 sigma
% A15  p-val(2,0.108)=0.0052 /   2.6 sigma
% A16  p-val(2,0.090)=0.0036 /   2.7 sigma

%%%    Ris_4
%%%
\begin{figure}[!htb]
%\begin{center}
\includegraphics[width=0.50\textwidth,height=0.35\textwidth]{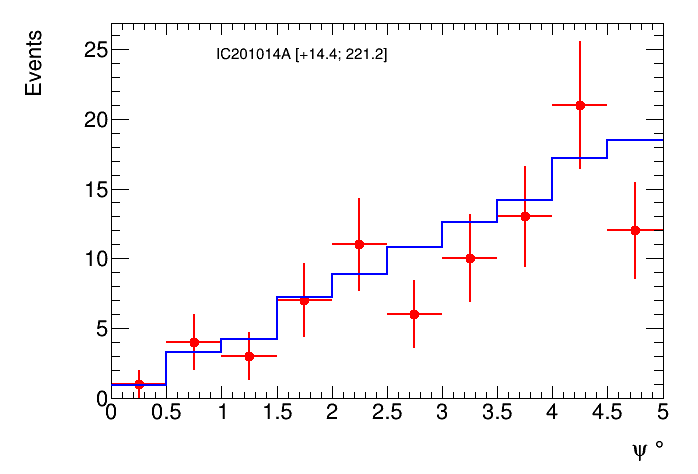}
\includegraphics[width=0.50\textwidth,height=0.35\textwidth]{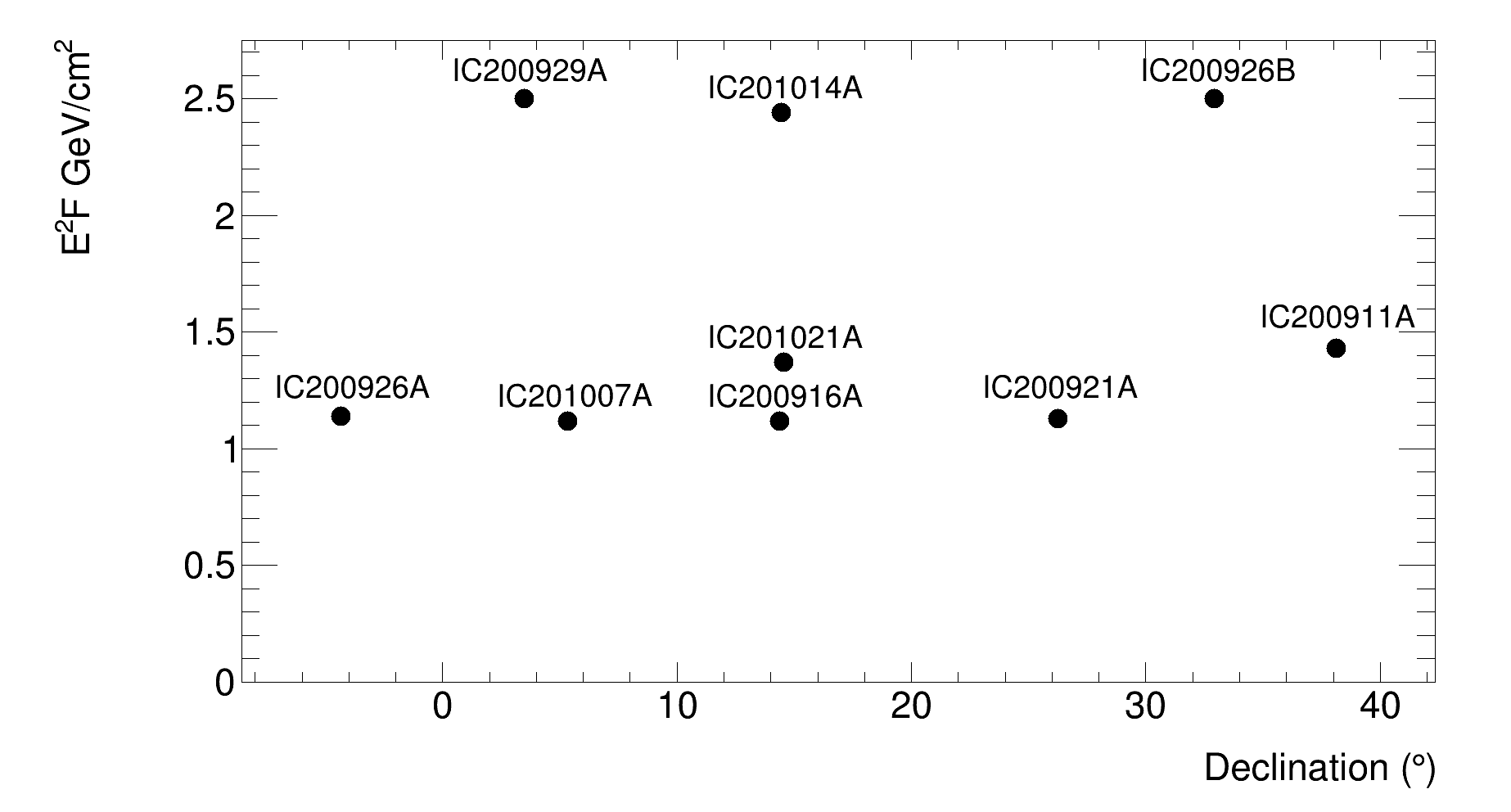}
%\includegraphics[width=0.50\textwidth,height=0.35\textwidth]{back_elev.png}
%\end{center}
\caption{Left: Distributions of numbers of observed cascades (red) and expected events around direction of alert IC201014A in October 14, 2020. 
Right: The upper limits on neutrino fluence towards IC astrotracks in 2020 (see text for details).
}
\label{fig-4}
\end{figure}
%%%
%%%
      The IceCube (IC) neutrino alerts classified as "astrotracks" are muons detected in a real-time with estimates on energy above 100 TeV and  
signal-to-background ratio above 30$\%$ ("bronze" event) or 50$\%$ ("gold"). Baikal-GVD is following up GCN/AMON messages on astrotracks 
since August 2020 with the beginning of the fast reconstruction analysis~\cite{AL2021}. Fig.2 (right) shows locations of 21 astrotracks by crosses and their duirnal visibilities in the GVD horizon coordinates. It is seen that most part of astrotracks are observed at the GVD as downgoing events or near horizon events with zenith angles less than 120$^\circ$. Thus, in prompt searches of event coincidences we used only cascades mode reconstruction. Two time windows of $\pm$1 hour and $\pm$12 hours have been taken in search for correlations with alert in a cone of 5${^\circ}$. Example of histograms of the observed and background number of events in a half-open cone of 5$^\circ$ 
are shown in Fig.4 (left) for alert IC201014A, where background is taken by the bootstrup method in mixing of real data sample for season 2020.
There were no events observed in the time window $\pm$1 hour around each alert, while for few of them there were a single cascades during larger time interval $\pm$12 hours. In our recent work in~\cite{AL2021} it was shown that a level of the expected backgound in the time window $\pm$12 hours varies between 0.29 and 0.45 events in cone 
of 5${^\circ}$ around alerts. Therefore, there is no any statistically significant excess of the number of observed events above the expected backgound. The upper limits at 90$\%$ confedence level (c.l.) on numer of events have been obtained for each IC alert according to Feldman and Cousins statistics~\cite{Feld}. Finally, constraints 
at 90$\%$ c.l. on the energy-dependent fluence of neutrinos of one type with a E$^{-2}$ spectrum under assumption of an equal fraction of neutrino flavors in the total fluece have been set for the IC alerts (see details in~\cite{AL2021}). In Fig.4 (right) one can see reached level of the Baikal-GVD sensitivity to astrophysical neutrino sources of the Northen sky in the shown example for 9 IC astrotracks as dependence of the upper limits on neutrino fluence versus declination values. 

\section{Follow-up of multi-wavelength signals}
 Most probable multimessenger signals might be observable towards a burst with associated radio emission. 
 The first HE alerts of the Baikal-GVD have been included in joint test of observing sample of the AGN of the 600~m RATAN radio telescope of the Special Astrophysical Observatory and the 40~m Telescope of the Owens Valley Radio Observatory (OVRO) as shown at the conference (Zh.Dzhilkibaev and I.Belolaptokov, PoS(ICRC2021)002. 
In our off-line analysis of the Baikal-GVD data samples we have followed up two interesting cosmic phenomena in April 2019 (TDE) and April 2020 (SGR). The fisrt one is the optical transient AT2019dsg discovered by the Zwicky Transient Facility (ZTE) 2019-04-09 at 11:09:28.0 UTC with coordinates: $\alpha=$ 314.262${^\circ}$ and $\delta= +$14.204${^\circ}$ and classified as a tidal disruption event~\cite{ZTE-AT2019dsg}. It is a known process when a star in a host galaxy with the supermassive black hole in its center can be gravitationally captured and torn apart by the black hole’s strong tidal forces. The TDE emits a bright flare of light, which then decays on timescales from months to years (see e.g.~\cite{EvansKochanek1989}) . Dedicated follow ups of the AT2019dsg at the neutrino telescopes IceCube and ANTARES claimed two neitrino events as most probable in correllation towards the TDE: 
IC191001A (1 October, 2019) and ANT200211 (11, February 2020)~\cite{IC-ANT-TDE}, that is after the first burtst at about 150 days and 250 days, accordinally.
We reported here our very preliminary results on a correlation analysis towards the AT2019dsg. Taking the cascades sample of a season 2019 with a live time of 323 days 
we search for events in a cone of 5$^\circ$ towards the AT2019dsg. Among events reconstructed at five clusters we have found 3 events at 3 different clusters with the delay time less than 12 hours between each other: MJD 58603.92444 (cluster 1), MJD 58603.82667 (cluster 3) and MJD 58603.95556 (cluster 4). Fig.5 shows events on each cluster that happened during a whole timescale of a season 2019 in a cone 5${^\circ}$. Each event is weighted accordingly to inverse value of mismatch angle between event and the source. Such way allows to see most closest events
to the source in relative difference of their time. On base of random sample of scrambling data we estimates the expected number of background events as 0.485 in the time interval $\pm$12 hours. That means a p-value to be 1.2$\%$ (2.26$\sigma$). There are two features in found triplet of cascades. One is that the found delay time after the first TDE-burst is 21 days and fixed day is April 30 of 2020. That what was predicted in some of theoretical models of the TDE-process (see~\cite{WinterLunardini}). Another important thing is that the time interval between these 3 cascades are much less than 12 hours: less than 1 hour for one pair and less than 3 hours between others. Further statistical analysis is in progress now.  
%%%$https://www.ovro.caltech.edu/$)
%%https://sites.astro.caltech.edu/ovroblazars/) 
% %and such catostrophical cosmic events as a busrt of TDE transients.
%%%
%%%
%%%    Ris_5
%%%
\begin{figure}[!htb]
%\begin{center}                                                                            
\includegraphics[width=0.33\textwidth,height=0.35\textwidth]{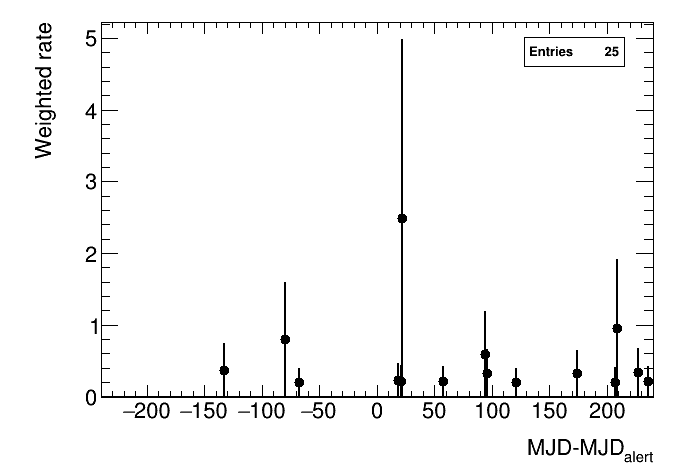}
\includegraphics[width=0.33\textwidth,height=0.35\textwidth]{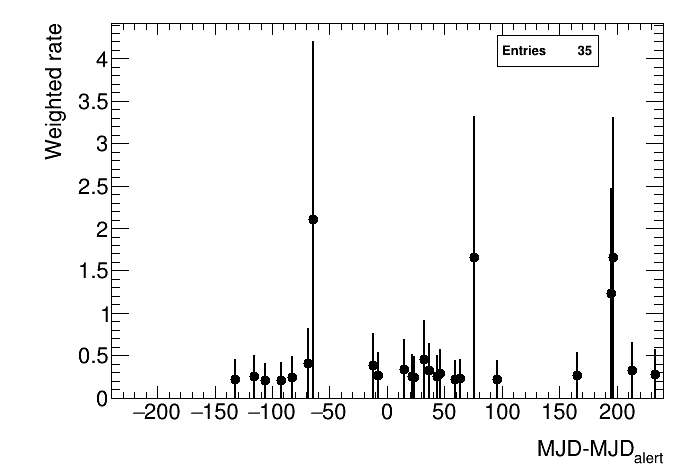}
\includegraphics[width=0.33\textwidth,height=0.35\textwidth]{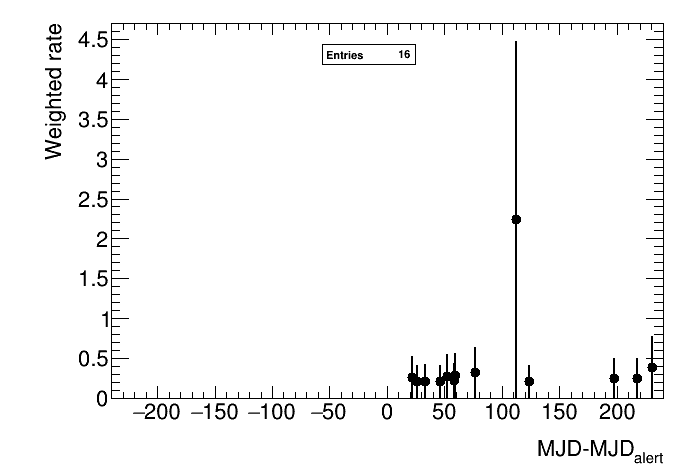}
%\end{center}
\caption{Weighed rates of cascade events each day at three GVD clusters (1, 3, 4) around the ATdsg2019 direction in a cone of 5$^\circ$ in time scale of relatively a discovery time of 
the optical burst.
}
\label{fig-5}
\end{figure}

The second promising source, that we have analysed in search for coincidences between it and the cascades sample of a season 2020, was the magnetar SGR 1935+2154 following to the INTEGRAL discovery of its new activity in period of radio and gamma burst in April 27-28 of 2020 and CHINE/FBR observed radio burst 28 April 2020 14:34:33 UTC ~\cite{INTEGRAL}. The magnetar SGR 1935+2154 lies in the Galactic Plane ($l=57.25^\circ,  b=+0.82^\circ$) and  could  be  associated with the supernova remnant G57.2+0.8 at distances less than 12.5 kpc. Our search method was close to one for the follow-up of neutrino alerts. With data of the first 100 days after the INTEGRAL alert we have analysed cascades distributed per clusters in few cones: less than 20, 15, 10, and 5 degrees; and two levels of hits rejection i.e. $N_{hits}>7$ and $_{Nhits}>9$. Finally, 2 cascades rests in 5 degrees around the source, satisfying both selections on hits. Expected background was estimated 
as 0.35 from analysis of real data sample of a season 2016 and 2018. Thus a p-value was found to be 4.87$\%$. The visibility of the SGR 1935+2154 in the Baikal-GVD horizontal coordinates 
has been calculated for integrated exposure. The upper limits at 90$\%$ c.l. on number of events was found to be 5.91. Finally, the upper limits at 90$^\%$ c.l. on the energy-dependent 
fluence of neutrinos of one type with a E$^{-2}$ spectrum under assumption of an equal fraction of neutrino flavors in the total fluece have been resulted in 2.0 GeV$\cdot$ cm$^{-2}$.  

In summary, the Baikal-GVD of 0.4 km$^3$ of effective volume for electron neutrinos fullfills complementary part of the sky observation and succesfully goes to a real-time alerts both in follow-up regime and catching high energy astrophysical neutrinos.

%\acknowledgments
%This work was supported by the Russian Foundation for Basic Research (Grants 20-02-00400, 19-29-11029).
The work was partially supported by RFBR grants 20-02-00400 and 19-29-11029. 
The CTU group acknowledges the support by European Regional Development Fund-Project No. CZ.02.1.01/0.0/0.0/16 019/0000766.
O.V. Suvorova acknowledges the supported by RSF grant no. 17-12-01547 in part of the works on data analysis.

%% Full authors list (ONLY FOR COLLABORATIONS)
%\clearpage
%\section*{Full Authors List: \Coll\ Collaboration}
%
%\noindent \textbf{Note comment afterwards:} Collaborations have the possibility to provide an authors list in xml format which will be used while generating the DOI entries making the full authors list searchable in databases like Inspire HEP. For instructions please go to icrc2021.desy.de/proceedings or contact us under icrc2021proc@desy.de.\\
%
%\scriptsize
%\noindent
%first.author$^1$, 
%second.author$^2$, 
%third.author$^3$ % .... more names
%and 
%last.author$^{n}$ \\
%
%\noindent
%$^1$first.affiliation.
%$^2$second.affiliation. % .... more affiliation
%$^{m}$last.affiliation.

\end{document}